\documentclass[12pt]{emulateapj}

\shorttitle{The Clustering of AGN in the SDSS}
\shortauthors{Wake et al.}

\begin{document}

\title{The Clustering of AGN in the Sloan Digital Sky Survey}
\author{David A. Wake$^1$, Christopher J. Miller$^1$,  
Tiziana Di Matteo$^2$, Robert C. Nichol$^1$, Adrian Pope$^3$, Alexander S. Szalay$^3$, Alexander Gray$^4$, Donald P. Schneider$^5$, Donald G. York$^6$}
\altaffiltext{1}{Dept. of Physics, Carnegie Mellon Univ.,
Pittsburgh, PA 15213}
\altaffiltext{2}{Max Planck Institute for Astrophysics Garching,
D-85741 Garching, Germany}
\altaffiltext{3}{Dept. of Physics and Astronomy, Johns Hopkins Univ., Baltimore, MD}
\altaffiltext{4}{Dept. of Computer Science, Carnegie Mellon Univ.,
Pittsburgh, PA 15213}
\altaffiltext{5}{Dept. of Astronomy and Astrophysics, Pennsylvania State Univ., University Park, PA 16802}
 \altaffiltext{6}{Dept. of Astronomy and Astrophysics and Enrico Fermi Institute, The University of Chicago,
Chicago, IL 60637}
\begin{abstract}
  We present the two--point correlation function (2PCF) of
  narrow-line active galactic nuclei (AGN) selected within the First
  Data Release of the Sloan Digital Sky Survey. Using a sample
  of 13605 AGN in the redshift range $0.055 < z < 0.2$, we find that
  the AGN auto--correlation function is consistent with the observed
  galaxy auto--correlation function on scales $0.2h^{-1}$Mpc to
  $>100h^{-1}$Mpc.
  The AGN hosts trace an intermediate
  population of galaxies and are not detected in either the bluest
  (youngest) disk--dominated galaxies or many of the reddest (oldest)
  galaxies.
  We show that the AGN 2PCF is
  dependent on the luminosity of the narrow [OIII] emission line
  ($L_{[OIII]}$), with  low $L_{[OIII]}$ AGN having a higher
  clustering amplitude than high $L_{[OIII]}$ AGN. This is
  consistent with lower activity AGN residing in more massive
  galaxies than higher activity AGN, and $L_{[OIII]}$ providing a
  good indicator of the fueling rate. Using a model relating halo mass
  to black hole mass in
  cosmological simulations, we show that AGN
  hosted by $\sim 10^{12}$ M$_{\odot}$ dark matter halos
  have a 2PCF that matches that of the observed sample.
  This mass scale implies a mean black hole mass for the
  sample of $M_{BH} \sim 10^8$ M$_{\odot}$.
\end{abstract}


\section{Introduction}

The clustering of galaxies as a function of their properties
provides important constraints on models of galaxy formation and
evolution. Such clustering is often measured using the two--point
auto--correlation function \citep[2PCF; see][]{1980lssu.book.....P}. In hierarchical models of structure formation, the amplitude of the
2PCF depends upon the mass of the dark matter halos \citep[{\it i.e.,}
more massive halos are clustered more
strongly;][]{1986MNRAS.222..323K}, while the shape of the 2PCF can
depend upon the details of how galaxies reside in those dark matter
halos \citep[]{2004zehavi}. For example, the amplitude and slope
of the 2PCF is lower for blue galaxies than for galaxies with the
reddest colors \citep[{\it e.g.},][]{1976ApJ...208...13D,2002ApJ...571..172Z}.

In this letter, we continue our study of the relation between the
environment of galaxies in the Sloan Digital Sky Survey
\citep[SDSS;][]{2000AJ....120.1579Y} and their observed physical
properties
\citep[see][]{2003ApJ...584..210G,2003ApJ...597..142M,2004balogh}. In
particular, we present the redshift--space 2PCF for a subset of SDSS
galaxies spectroscopically classified as narrow-line active
galactic nuclei \citep[AGN;][]{2003ApJ...597..142M}. Our analysis
has two advantages over previous measurements of the AGN 2PCF: larger
sample size (in number and area), and a homogeneous selection criteria
(compare to Table 1 of \citet{2001AJ....122...26B} for previous
AGN 2PCF measurements). In addition, the data are now large enough
to study both volume-limited subsamples as well as how AGN or AGN
host galaxy properties affect the 2PCF.

\section{Data}
\label{data}

We use the main galaxy sample data \citep{2002AJ....124.1810S}
from the First Data Release (DR1) of the SDSS
\citep[see][]{2003AJ....126.2081A}. To select AGN, 
we have used the methodology presented in
Section 2.1 of \citet{2003ApJ...597..142M}, where the
AGN are classified using the emission-line flux ratios 
log([OIII]/H$\beta$) versus log([NII]/H$\alpha$) 
\citep[see][]{2001ApJS..132...37K} or simply 
log([NII]/H$\alpha$)$> -0.2$, if [OIII] or H$\beta$ are not measured \citep[see
also][]{2001ApJ...559..606C,2004brinchmann}. 
We remove all galaxies from areas with high seeing values
($>$ 2 \arcsec) and $r$ band Galactic extinction $> 0.4$ magnitudes. These
restrictions produce a sample of 72455 SDSS DR1 galaxies within $0.055 \le z \le 0.2$, from which
we classify 13605 galaxies as AGN. This fraction of AGN
(18\%) is consistent with the findings of \citet{2003ApJ...597..142M} and 
\citet{2004brinchmann}.
We discuss implications of our
classifications in Section \ref{sec::properties}.

\section{AGN and Galaxy Correlation Functions}

We account for the survey geometry (or
mask) by constructing random catalogs
that match both the survey angular and radial selection functions. We
first construct a random catalog that has the same angular mask as the
real data.
We then construct the radial selection function
by smoothing the observed
redshift distributions
with a Gaussian of width $z = 0.025$. These
smoothed redshift distributions are used to randomly assign
redshifts to the data points in our random catalogs, which are
ten times larger than the real datasets. We note that our conclusions
are robust to the choice of width for the smoothing kernel.
We calculate the 2PCF using the \citet{1993ApJ...412...64L} estimator
and estimate the covariance using the
jack-knife resampling technique
\citep[{e.g.}][]{1993stp..book.....L,2002ApJ...571..172Z} splitting the 
angular mask of our data into 32 subsections of $\simeq
10^{\circ}\, \times\, 5^{\circ}$ (or $\simeq 60\, \times\, 30\,
h^{-1}$ Mpc at $z = 0.1$).
We use $H_0=100{\rm
km\,s^{-1}\,Mpc^{-1}}$, $\Omega_{m}=0.3$, and $\Omega_{\Lambda}=0.7$.

\begin{figure}[t]
\plotone{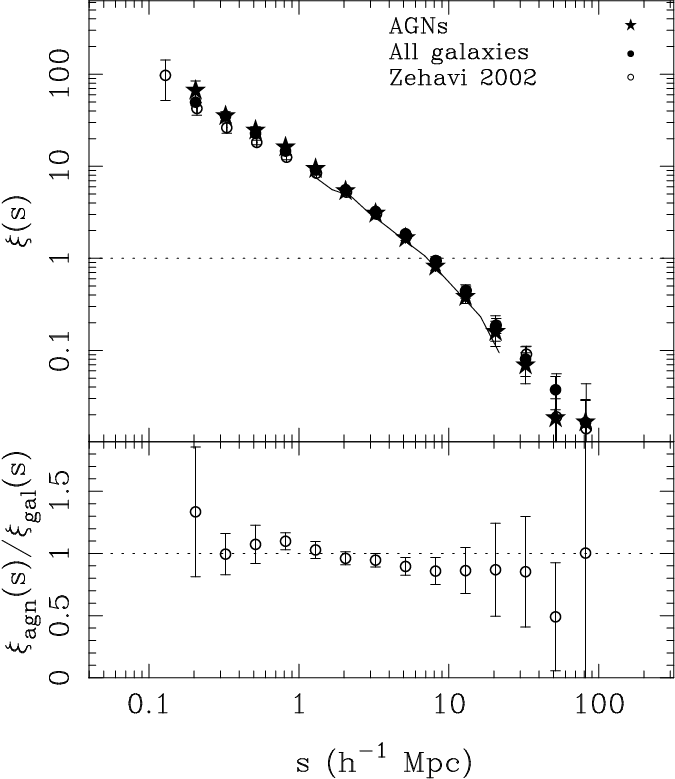}
\caption{(Top) The 2PCF for AGN and all SDSS galaxies. We also show the 2PCF
of Zehavi et al. (2002). 
The solid line is the model from Section \ref{sims}.
(Bottom) The ratio of the AGN and
galaxy 2PCFs. 
 \label{twopoint}}
\end{figure}

In Figure \ref{twopoint}, we show the redshift--space 2PCF for both the
AGN (stars) and all galaxies (filled dots) samples discussed in Section \ref{data}. 
We also show
the SDSS redshift-space 2PCF of \citet{2002ApJ...571..172Z}. As
expected, our galaxy 2PCF agrees with \citet{2002ApJ...571..172Z},
except on small scales due to incompleteness from fiber collisions
\citep{2003AJ....125.2276B}, since Zehavi et al. attempt to correct
for these collisions. 
We note that since the fiber collisions are uniformly distributed
over the survey (see Blanton et al. 2003a), they affect both the galaxy and AGN 2PCF in the same
way. Thus we study only relative differences between 2PCFs here.
We measure a simple $\chi^2$ statistic between the two samples and account for errors on both 2PCFs by combining their
individual covariances. Specifically, we take
the square root of the sum of the squared covariances, which accounts
for correlations between data points of
different separations in both datasets.
We find $\chi^2=13$ with $9$
degrees-of-freedom for pair separations greater than $\sim
1h^{-1}$Mpc, {\it i.e.}, there is no significant difference between
the AGN and galaxy 2PCF.
We show the ratio of the two 2PCFs
in the bottom of Figure \ref{twopoint}.
The weighted mean ratio is
$\xi_{agn}/\xi_{gal}=0.974\pm0.026$.

\begin{figure}
\plotone{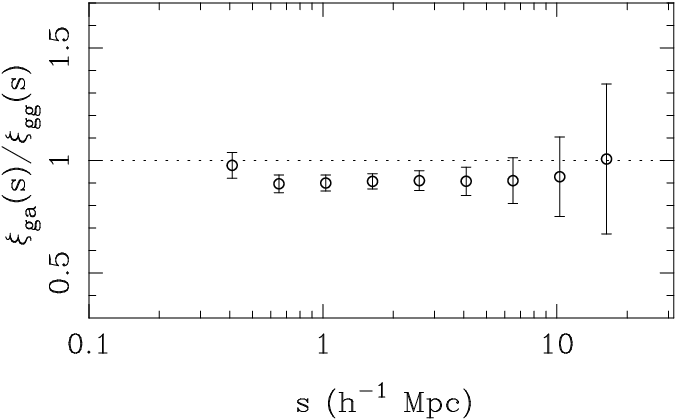}
\caption{The ratio of the AGN--galaxy cross--correlation function to the
galaxy--galaxy auto--correlation function. 
\label{cross}}
\end{figure}

We show in Figure \ref{cross} the ratio of the AGN--galaxy
cross--correlation function \citep[see also ][]{1999MNRAS.305..547C,2003croom} to
the normal galaxy 2PCF again for the samples defined in Section \ref{data}. The weighted mean ratio between these two
functions is $\xi_{agn-gal}/\xi_{gal-gal}=0.922\pm 0.028$. Within the one sigma uncertainties,
this is consistent with
\citet{2003croom} who demonstrate that, for $z<0.3$ quasars, the
ratio of the quasar--galaxy cross--correlation function is
$\xi_{QSO-gal}/\xi_{gal-gal}=0.97\pm0.05$. 

Finally, we measure the 2PCF as a function of AGN activity as measured
by the luminosity of the forbidden [OIII] narrow emission line
($L_{[OIII]}$). This line is proposed to be only weakly affected by any
residual star--formation \citep[see][]{2003MNRAS.346.1055K}. However, 
the exact connection
between the [OIII] strength and AGN activity is not well established
in narrow-line systems
\citep[see][]{2000ApJ...544L..91N,2002ApJ...565...78B,2000MNRAS.314L..17M,2001NewA....6..321M,2004grupe}.
We created two sub-samples of AGN, one containing the highest third
of the $L_{[OIII]}$ distribution ($>$ 4.84 $\times$10$^{44}$ ergs$^{-1}$) and one with the lowest third ($<$ 1.29
$\times$10$^{44}$ ergs$^{-1}$). In order to minimize any selection
bias ({\it e.g.} Malmquist bias), we construct a pseudo-volume
limited sample by restricting the AGN sample to a redshift range of
$0.06 < z < 0.085$ and to a k-corrected \citep{2003AJ....125.2348B}
absolute magnitude limit of M$_r < -19.8$
\citep[see][]{2003ApJ...584..210G,2004balogh}. This provides a sample
of 2457 AGN. We find that
the distributions of host galaxy absolute magnitudes and redshifts are identical for
the low and high $L_{[OIII]}$ samples and to the entire galaxy sample.

\begin{figure}
\plotone{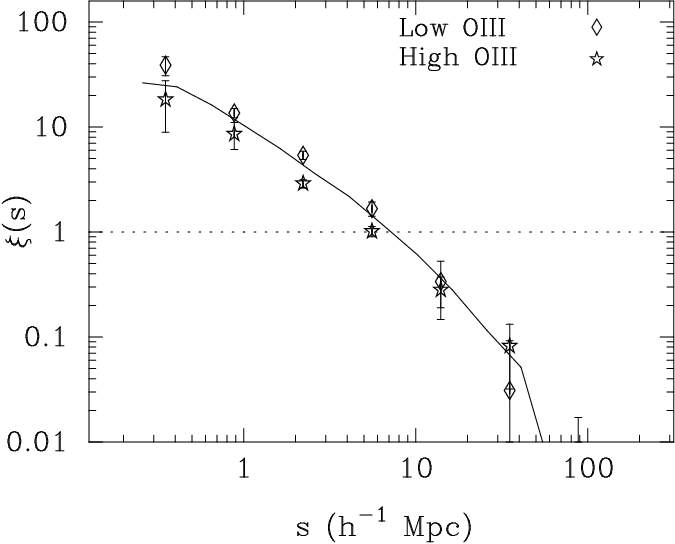}
\caption{The AGN 2PCF as a function of $L_{[OIII]}$. We show the 2PCF for 
AGN in the top and bottom third of the
$L_{[OIII]}$ distribution. The solid line is for all SDSS galaxies taken
from Figure \ref{twopoint}.
\label{hilo_o3}}
\end{figure}

In Figure \ref{hilo_o3}, we present the AGN 2PCF as a function of
$L_{[OIII]}$. We see a noticeable difference in the amplitude of
clustering, with the lower luminosity AGN having a stronger
clustering amplitude. We calculate the $\chi^2$
difference between the high and low $L_{[OIII]}$ sub-samples and
find $\chi^2=40$ with $5$ degrees-of-freedom. Therefore, the 2PCFs
for the high and low $L_{[OIII]}$ sub-samples are different at
the $> 5\sigma$ level. We also see a similar difference in the
clustering amplitudes if we split the AGN sample as a function of the
width of the [OIII] emission line.

\begin{figure}
\plotone{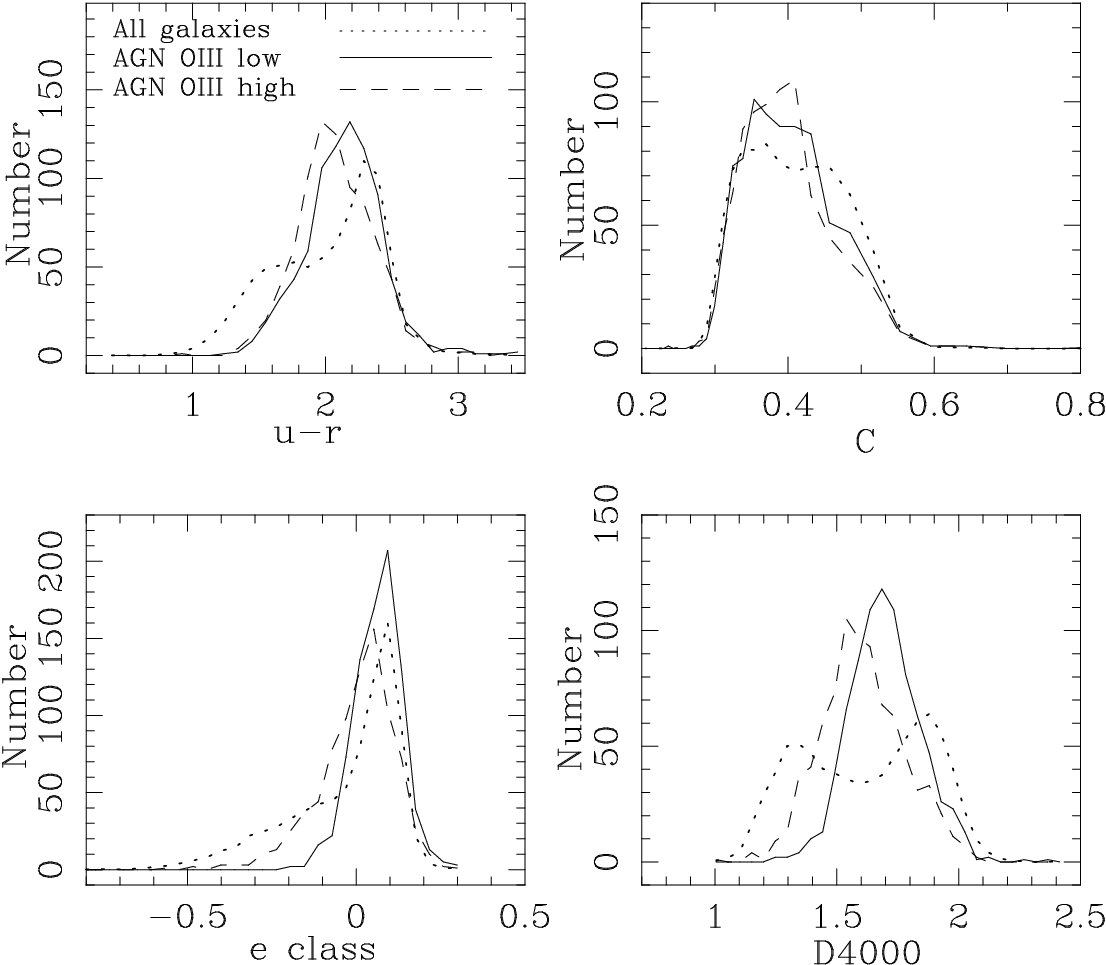}
\caption{The distributions of galaxy properties (described in the text) for our volume--limited sample.
All distributions
have been re--normalized to have the same total number.
\label{distributions}
}
\end{figure}

\section{Discussion}

\subsection{The Properties of AGN Host Galaxies}
\label{sec::properties}

Throughout this paper we assume that the measured galaxy properties reflect those of the host
galaxies and are not significantly affected by the AGN light. Both \citet{1999MNRAS.303..173S}
and \citet{2003MNRAS.346.1055K} find that the AGN contribution to the total luminosity 
in narrow-line AGN like those studied here rarely exceeds 5\%. We measure the AGN contribution
using the ratio of the total flux over the flux within the 3\arcsec~ fiber
and find that the AGN contribute on average $<$ 6\% of the total light.

Using our magnitude- and volume-limited sample, we find that the distributions of the AGN host
galaxy properties are
different from those of all galaxies (Figure \ref{distributions}). For example, the concentration index (C) \citep{2001AJ....122.1238S} and
e-class \citep{2002AJ....123..485S,1998IAUS..179..376C} for our AGN
sample appear to be preferentially missing the bluest (in e-class),
disk--dominated galaxies \citep[see also Figure 12
in][]{2003ApJ...597..142M}. 
The most striking difference 
is for the D4000 index, a measure of the 4000\AA~ break, and the $u-r$ color, where the AGN distributions 
do not exhibit the bi-modal shape seen for all galaxies.

One possible explanation for the difference between the AGN and all galaxy distributions
is our exclusion of the broad-line QSOs.
To investigate this, we have used the $94$
broad-line SDSS QSOs within $0.06 \le z \le 0.085$ and brighter than M$_{r}$ = -19.8.
These QSO host galaxies have a wide
variety of morphologies, while 20\%
lack an obvious host galaxy ({\it i.e.}, they appear point-like in the SDSS), and so any
k-corrections on this population could be inaccurate. However, the observed-frame
colors of all of the QSOs are on average bluer than our AGN sample,
as expected if the QSO component dominates over the light from the host galaxy.
Even so,
the total number of these broad-line QSOs is simply too small to accommodate
the missing blue host galaxy populations discussed above. 

Another possibility comes from our AGN selection
criteria. 
In particular, our signal--to--noise limit on
the SDSS spectra could preferentially exclude the lowest luminosity AGN, as
they could be either buried in strongly star--forming bulges or accreting at a
very low rate in the reddest, oldest galaxies. Dust obscuration
could also significantly affect our detection of AGN, especially for the
strongly star-forming (bluest) galaxies \citep{2003ApJ...599..971H}.
For instance, $\sim 30\%$ of our galaxies have emission-lines but could not 
be classified as either star-forming or AGN, and could
be obscured AGN. 
\citet[]{2003ApJ...597..142M} attempted to statistically model
this population using colors and noted that there was a significant
red population, which were most likely AGN. Using this model, we have
classified these unidentified emission-line galaxies (ELUs) as AGN.
This has the effect of adding mainly red (but some blue) galaxies to the histograms in Figure
\ref{distributions} although they are still dominated by galaxies intermediate between blue and red (low and high D4000).
We recalculate the 2PCF including these model-dependent AGN classifications, and find no statistical difference from the AGN 2PCF
that ignores the ELUs.
We have not attempted to subtract off the stellar components of the SDSS spectra \citep[see {\it e.g.},][]{2004cbhg.sympE..23H}, and
so we cannot rigorously address those galaxies which have both star-forming and AGN components. However,
as noted in \citet{2003ApJ...597..142M}, the fraction of late-type (e.g., spiral) morphologically classified galaxies harboring an AGN is
$\sim 20\%$, which is similar to that found by \citet{2004cbhg.symp..293H} who does subtract off stellar templates. The consistency
between \citet{2004cbhg.symp..293H} and \citet{2003ApJ...597..142M} suggests that there are not large numbers of AGN in strongly star-forming
galaxies that we fail to detect.

In summary the exclusion of the QSOs does
not explain why our AGN sample is lacking the bi-modal color
distribution of the whole galaxy sample. Likewise, while we are
certainly missing some AGN in the unidentified emission-line
galaxy population, our measured 2PCF is not altered after we attempt
to include them.
These issues could be addressed in a more detailed way through a multi--wavelength study of
these unidentified emission-line objects. 
Therefore, as suggested in \citet[]{2003ApJ...597..142M}, our AGN sample appears
to be an unbiased tracer, with respect to mass of the whole galaxy population,
for the large-scale structure in the local Universe. 

Given that the typical AGN host properties are not a random sub-sample of all
galaxies, it is somewhat surprising that they should cluster the same
way as all galaxies. For example, it has been shown that the 2PCF is a
strong function of both the color and luminosity of galaxies
\citep{1976ApJ...208...13D,1988ApJ...331L..59H,2002ApJ...571..172Z}, indicating that
the bluest, youngest galaxies preferentially populate the lowest
density regions in the Universe,
while the reddest, oldest
galaxies preferentially live in the densest regions \citep[{\it e.g.},][]{1974ApJ...194....1O,1980ApJ...236..351D,2003ApJ...584..210G,2004balogh}. Removing these two tails of the distribution could
result in an intermediate population that clusters the same way as the whole sample. 
Assuming that the existence of an AGN is independent of environment 
\citep[see ][]{2003ApJ...597..142M}, one can conclude
that the mean mass of the AGN dark matter halos must be the same
as the mean for all galaxies (see next Section). 


\subsection{The Mass Scale Selected by the AGN}
\label{sims}

We have used the model of \citet{2003ApJ...593...56D} to
make a prediction for the AGN 2PCF. In this model,
cosmological hydrodynamical simulations
\citep{2003MNRAS.339..289S,2003MNRAS.339..312S} were used to link the
growth and activity of central black holes in galaxies to the
formation of spheroids in galaxy halos. 
In the prescription used
for black hole growth in the simulations, it is assumed that the black
hole fueling rate is regulated by star formation in the gas. This
simple assumption was shown to explain both the observed $M - \sigma$
relation \citep{2000ApJ...539L...9F,2000ApJ...539L..13G} and the broad
properties of the AGN luminosity function (for an assumed quasar
lifetime). We use this model to construct a mock AGN catalog
and deduce that the minimum dark matter halo mass ($M_{\rm min}$) in
the simulations that best matches the observed space density of SDSS
AGN ($n \sim 1.5 \times 10^{-3}$ Mpc$^{-3}$) is $M_{\rm min} \ge 2
\times 10^{12}$ M$_{\odot}$. This is representative of low redshift
$L^*$ galaxies, which provide the bulk of our AGN sample.
Not surprisingly, the 2PCF for simulated dark matter halos with masses greater than
$M_{\rm min}$ (shown in Figure \ref{twopoint}) agrees well with
the observed 2PCF (also based on $\sim L^*$ and brighter galaxies). 
As mentioned in the last section,
we expect the AGN clustering amplitude to match that of the entire galaxy
population when the mean masses of the two populations are similar.

By relating the dark matter halo to the black hole mass
(according to Eqn. 8 of Di Matteo et al. 2003b; see also the observed
correlation by Ferrarese 2001 and Baes et al. 2003), we deduce a
mean black hole mass of our sample of AGN of $M_{BH} \sim 10^8$
M$_{\odot}$.

\subsection{Clustering and AGN Activity}

In Figure \ref{hilo_o3}, we show that the 2PCF for the lowest $L_{[OIII]}$
 AGN in our sample appears to have a higher clustering
amplitude than the highest $L_{[OIII]}$ AGN. We test whether this amplitude
difference is a result of the differing AGN host galaxy distributions.
 From the full AGN sample we randomly construct two sub-samples 
that possess the same D4000 distributions as shown in Figure
\ref{distributions} for the low and high $L_{[OIII]}$ samples,
but with no regard to the $L_{[OIII]}$. We find no difference
in their respective 2PCFs. We repeat the test for $u-r$ color, e-class and
concentration index and again find no difference. These tests
demonstrate that the difference seen in the clustering strengths
between the low and high $L_{[OIII]}$ samples is driven by the
[OIII] emission line and not by the underlying galaxy properties in
Figure \ref{distributions}. As an additional test, we split our AGN sample
into highest and lowest thirds of their D4000s, $u-r$ color, e-class, and concentration indices,
regardless of $L_{[OIII]}$.
In most cases, we do see differences in the 2PCF
of these subsamples {\it e.g.} the high D4000 sample is more strongly
clustered than the low D4000 AGN sample. Likewise, the redder AGN are
more clustered than the bluer AGN. However, the difference in clustering amplitude
is strongest when the AGN are split by
$L_{[OIII]}$. 

In hierarchical models of structure formation, more massive dark
matter halos are more strongly clustered. Therefore, the fact that the
low $L_{[OIII]}$ AGN sub-sample has a higher clustering amplitude
indicates that the host dark matter haloes of these AGN must be preferentially
more massive than the high $L_{[OIII]}$ AGN. Furthermore, $L_{[OIII]}$ delineates
 the high and low mass halos better than do other host galaxy properties (like D4000 or color).
If, as expected, the mass of the black hole correlates with the halo mass, then the weaker
 $L_{[OIII]}$ AGN must have larger black holes; therefore, a low $L_{[OIII]}$ can only be caused by a low fueling rate.
These observations are in accordance with studies of nearby massive
ellipticals, which are known to host the largest black holes
\citep[consistent with the $M-\sigma$
relation;][]{2000ApJ...539L..13G,2000ApJ...539L...9F}, but which
typically display the weakest AGN \citep{1997ApJ...487..568H,1999MNRAS.305..492D,2003ApJ...582..133D}.
Conversely, the high $L_{[OIII]}$ AGN have a lower clustering
amplitude consistent with them occupying lower mass dark matter halos
(hence having smaller central black holes) but accreting at a high
rate.

\acknowledgments

We thank Volker Springel and Lars Hernquist for their
Hydro-simulations performed at the Center for Parallel Astrophysical
Computing at the Harvard-Smithsonian Center for Astrophysics, Jon Loveday, 
David Weinberg, Michael Vogeley and the anonymous referee for useful discussions 
or comments. Funding
for the creation and distribution of the SDSS Archive has been
provided by the Alfred P. Sloan Foundation, the Participating
Institutions, the National Aeronautics and Space Administration, the
National Science Foundation, the U.S. Department of Energy, the
Japanese Monbukagakusho, and the Max Planck Society. The SDSS Web site
is http://www.sdss.org/.

\end{document}